RESEARCH ARTICLE									OPEN ACCESS

# Guardians of Trust: Navigating Data Security in AIOps through Vendor Partnerships

Subhadip Kumar
Western Governors University

**ABSTRACT**
Artificial Intelligence for IT Operations (AIOps) is a rapidly growing field that applies artificial intelligence and machine learning to automate and optimize IT operations. AIOps vendors provide services that ingest end-to-end logs, traces, and metrics to offer a full stack observability of IT systems. However, these data sources may contain sensitive information such as internal IP addresses, hostnames, HTTP headers, SQLs, method/argument return values, URLs, personal identifiable information (PII), or confidential business data. Therefore, data security is a crucial concern when working with AIOps vendors. This article discussed about the security features offered by different vendors and how best practices can be adopted to ensure data protection and privacy.

## I. INTRODUCTION

AIOps, or Artificial Intelligence for IT Operations, is a new approach that leverages machine learning and automation to enhance the observability and reliability of complex software systems. Observability is the ability to monitor and understand the internal state of a system or application based on the external outputs, such as logs, metrics, and traces. Several vendors offer full stack observability, monitor security vulnerabilities, automations, actionable alerts, and insights. These vendors ingest customers' logs, traces and metrices to produce actionable insights and recommendations that help SREs, and developers achieve full stack observability and improve the quality and efficiency of their software delivery. Security is a big concern when these vendors ingest customer data such as logs, metrices and traces as they consist sensitive information such as IP addresses, client details and their personal information's even confidential business data. Also, when this data is in transit or at rest, vendors ensure that the data is protected from being exposed. Security is always a joint responsibility that requires the participation and contribution of all stakeholders to create a safe and secure environment for everyone. In this article, we will discuss about different AIOps vendors and their security features. However, AIOps also poses some security challenges, such as data privacy, access control, and compliance. Therefore, it is important to choose an AIOps vendor that can provide robust and reliable security solutions. Different AIOps vendors have different standards and approaches to meet security standards. In this article, we will discuss in detail about those security measures and how a customer can leverage them and develop a best practice.

We will also discuss about how the vendors ingest customer data, encryption in transit and in rest from customers' system to vendor, in product communications, masking of sensitive information's and PII. Will also discuss how to protect the data using RBAC (Role based access control) and masking to ensure minimum exposure in case of a data breach.

## II. TYPE OF DATA COLLECTED

AIOps can help IT teams monitor, analyze, and troubleshoot complex systems, as well as improve service quality and customer satisfaction. However, AIOps also involves collecting and processing logs, metrics and traces which contains a large amount of sensitive data, such as client IP addresses, HTTP headers, HTTP post parameters, URL query parameters, SQL bind variables, SQL statements, personally identifiable information (PII), and more. This data can reveal a lot of information about the users, their behavior, their preferences, and their identity. If this data is not protected properly, it can lead to a serious data breach, which can have legal, financial, and reputational consequences for both the IT service provider and the users. Therefore, it is essential to implement appropriate security measures to protect the data at every stage of its lifecycle, from capture, to transit, to storage, to display.

## III. DATA COLLECTION AGENT BY VENDOR AND AGENT SECURITY

### A. DATA COLLECTION AGENT BY VENDOR

Most of the vendors deploy a single agent to collect data from remote sources and forward that to vendor's instance. Some examples include:
- Splunk has three types of forwarders aka agent – Universal forwarder, heavy forwarder, and light forwarder [1]. Out of that universal forwarder is the most popular one. Universal forwarder handles all kinds of data – starting from Microsoft Windows event logs, webserver logs, change logs, archive files etc.

- Dynatrace uses OneAgent – a single agent per host that collects all relevant metrics along with 100% of your application-delivery chain.
- AppDynamics uses different agents by application type – Java/.Net/Python/SAP are few of them. Agent collects data and forward it to the controller host for further processing.
- Datadog agent is deployed in OS is open source, and it is available on Github to consume by the user. It collects events and metrices from hosts and sends them to DogStatsD which is a metrics aggregator using StatsD protocol. DogStatsD also accepts custom metrics and events and forward them to DataDog.

### B. AGENT SECURITY

Agent security is an important aspect of AIOps, as agents are used by various vendors to collect and transmit sensitive data over the network. In this section, we will explore the different dimensions of securing the agents.

Agent distribution: The customer should ensure that the agents' repositories and binary packages are signed. They can validate the distribution channel by checking the signature against a public key.

Firewall rules: The customer should consult with the vendor to determine which inbound and outbound ports need to be opened for AIOps. The vendor documentation usually provides clear information on the ports that need to be open and the IP addresses that need to be whitelisted.

Information security and encryption: Data encryption in transit is another crucial aspect of agent security. The customer should verify that their agents use TLS-encrypted TCP connections and comply with the minimum TLS version required by their enterprise security team.

Agent runtime user: The customer should set the privilege and permission for the agent's runtime user as low as possible and only allow it to perform the assigned task.

### C. DATA STORAGE

AIOps vendors typically store customer data in secure and scalable storage environments, and the choice of storage solution may vary depending on several factors, including the vendor's infrastructure, specific use cases, and compliance requirements. Here are some common storage options that AIOps vendors may employ for storing customer data:

1) Cloud Storage Services:

Many AIOps vendors leverage cloud storage services provided by major cloud providers such as Amazon Web Services (AWS), Microsoft Azure, or Google Cloud Platform (GCP). These services offer scalable, reliable, and secure storage solutions with built-in encryption and access control features.

2) Databases:

AIOps vendors often use databases to store structured customer data. Relational databases (e.g., MySQL, PostgreSQL) or NoSQL databases (e.g., MongoDB, Cassandra) can be employed based on the nature of the data and specific requirements. Databases provide efficient data retrieval and management capabilities. For example, Dynatrace uses Grail, a database designed specifically for observability and security data. Grail is based on Dynatrace Query Language (DQL), which allows users to access and process data from different sources, such as logs, metrics, traces, events, and more. Grail also connects all the data within a real-time model that reflects the topology and dependencies within a monitored environment. This enables Dynatrace to provide a holistic view, advanced analytics, and AI-powered answers for cloud optimization and troubleshooting.

3) Object Storage:

Object storage is suitable for storing unstructured data like documents, images, and log files. It offers scalable and cost-effective storage solutions, and popular object storage services include Amazon S3, Azure Blob Storage, and Google Cloud Storage. Datadog often supports multi-cloud environments. They may store customer data in cloud storage solutions like AWS S3, Google Cloud Storage, or Azure Blob Storage.

4) On-Premises Storage:

Some AIOps vendors may opt for on-premises storage solutions, especially in cases where regulatory or compliance requirements mandate keeping data within a specific physical location. On-premises storage allows vendors to have more control over their infrastructure and data.

5) Hybrid Storage Solutions:

A hybrid storage approach combines on-premises and cloud-based storage solutions. This provides flexibility and allows vendors to optimize their storage strategy based on the specific needs of their AIOps platform, balancing performance, scalability, and compliance.

6) Distributed File Systems:

AIOps vendors dealing with large-scale data processing may utilize distributed file systems such as Apache Hadoop Distributed File System (HDFS) or distributed storage systems like Ceph. These systems are designed to handle massive amounts of data across distributed clusters.

## IV. DATA COLLECTION AGENT BY VENDOR AND AGENT SECURITY

### A. PREVENT CAPTURING OF SENSITIVE DATA

Sensitive data exposure is a serious risk for AI tools, especially those that use generative models to create content based on user inputs. Generative models are a type of machine learning models that can learn from data and generate new data that resembles the original data. For example, generative models can create realistic images, texts, or sounds based on user inputs. However, generative models can also inadvertently capture and expose sensitive data from the inputs, such as personal information, intellectual property, trade secrets, or confidential records. This can happen if the inputs contain such data, or if the models are trained on datasets that contain such data. If such data is leaked, stolen, or misused, it can cause reputational damage, legal liability, or competitive disadvantage for the data owners. Therefore, it is important to prevent capturing sensitive data right at the source before it is processed by the AI tool. One way to do this is to use data anonymization techniques, such as masking, hashing, encryption, or tokenization, to remove or replace any identifying information from the data. Masking is a technique that hides or obscures sensitive data with random characters or symbols. Hashing is a technique that transforms sensitive data into a fixed-length string of characters using a mathematical function. Encryption is a technique that converts sensitive data into an unreadable format using a secret key. Tokenization is a technique that replaces sensitive data with a unique identifier that maps to the original data in a secure database. These techniques can help preserve the utility and structure of the data, but without revealing any sensitive details. Another way to prevent capturing sensitive data at the source is to use data minimization techniques, such as filtering, sampling, or aggregation, to reduce the amount and granularity of the data. Filtering is a technique that removes or excludes data that is irrelevant, redundant, or sensitive from the inputs. Sampling is a technique that selects a subset of data that is representative of the whole population. Aggregation is a technique that combines or summarizes data into groups or categories. These techniques can help reduce the complexity and size of the data, but without containing any unnecessary or excessive information. By applying these techniques, data owners can protect their sensitive data from exposure and ensure the privacy and security of their data while using AI tools.

### B. MASKING DATA AT CAPTURE

Masking is a data anonymization technique that hides or obscures sensitive data with random characters or symbols. For example, you can replace a value character with a symbol such as "*" or "x". Masking can help preserve the utility and structure of the data, but without revealing any sensitive details.

Masking works by applying a masking function to the sensitive data at the source, before it is processed by the AI tool. The masking function can be deterministic or random, depending on the level of security and consistency required. A deterministic masking function always produces the same output for the same input, while a random masking function produces different outputs for the same input. A deterministic masking function can ensure that the masked data is consistent across different datasets, while a random masking function can increase the difficulty of reverse engineering the original data.

Masking is useful for protecting sensitive data from unauthorized access or disclosure, while still allowing the data to be used for certain purposes, such as testing, training, or analysis. Masking can also reduce the risk of data breaches, as the masked data is less valuable or attractive to hackers or malicious insiders. Masking can also help comply with data privacy regulations, such as GDPR, that require data minimization and pseudonymization.

However, masking also has some challenges or limitations, such as:

- Masking can reduce the quality or accuracy of the data, as some information is lost or distorted during the masking process. This can affect the performance or reliability of the AI tool that uses the masked data.

- Masking can be vulnerable to re-identification attacks, especially if the masked data is combined with other data sources that can reveal the identity of the data subjects. For example, if the masked data contains a unique identifier, such as an email address, that can be linked to another dataset that contains the name of the data subject, the masked data can be re-identified.

- Masking can be difficult to implement or maintain, especially if the data is dynamic or complex. For example, if the data changes frequently or has multiple formats or types, the masking function may need to be updated or customized accordingly. This can increase the cost and complexity of the data anonymization process.

Therefore, it is important to follow some best practices for masking, such as:

- Designate a multi-tier access and authorization system for your most critical assets. Whenever it's necessary for larger groups of people to access or use sensitive data, look for additional ways to protect that data, such as data encryption, anonymization, and/or masking.

- Choose the appropriate masking function and level for your data, depending on the level of security and consistency required. For example, you can use deterministic masking for data that needs to be consistent across different datasets, or random masking for data that needs to be more secure.

- Test and validate the masked data to ensure that it meets the quality and usability requirements for the AIOps tool. For example, you can use data quality metrics, such as completeness, correctness, and consistency, to measure the quality of the masked data. You can also use data usability metrics, such as utility, relevance, and timeliness, to measure the usability of the masked data.

### C. MASKING AT INGEST

Another way to mask the sensitive data once it reaches the AIOps SaaS platform is to use masking at ingest. This approach allows masking sensitive data once it arrives in the SaaS environment, and before it is written to disk (stored). This way, the data can still be used for the intended purpose, but without revealing any sensitive details.

Masking at ingest works by applying a masking function to the sensitive data as soon as it is received by the AIOps SaaS platform. The masking function can be configured by the user, based on the type and level of sensitivity of the data. For example, the user can choose to mask all or some of the fields in a log entry, such as IP address, username, or email address. The user can also choose the masking method, such as replacing, hashing, or encrypting the sensitive data. The masked data is then stored in the AIOps SaaS platform, while the original data is discarded or archived.

### D. TLS/SSL ENCRYPTION WHILE DATA AT TRANSIT

One way to protect the data in transit is to use TLE/SSL encryption, which stands for Transport Layer Encryption/Secure Sockets Layer encryption. TLE/SSL encryption is a protocol that provides security and privacy for network communications. TLE/SSL encryption works by encrypting the data before transmission, authenticating the endpoints, and decrypting and verifying the data on arrival. Encryption means converting the data into an unreadable format using a secret key, which prevents anyone from accessing or modifying the data without the key. Authentication means verifying the identity and trustworthiness of the endpoints, which prevents anyone from impersonating or intercepting the communication. Decryption means converting the data back into a readable format using the same or a different key, which ensures that the data is intact and has not been tampered with.

TLE/SSL encryption can enhance the security and privacy of the data in transit by preventing unauthorized access, disclosure, or modification of the data. According to a study by Krawczyk et al. [3], TLE/SSL encryption can provide a strong security notion for network communications, called authenticated and confidential channel establishment (ACCE), which guarantees that the data is protected from both passive and active attacks, such as eavesdropping, tampering, replaying, or forging. TLE/SSL encryption can also help comply with data privacy regulations, such as GDPR, that require data protection and confidentiality. According to a report by ENISA [2], TLE/SSL encryption can help data controllers and processors meet the GDPR requirements for data security, such as ensuring the confidentiality, integrity, and availability of the data, as well as the resilience of the systems and services that process the data. TLE/SSL encryption can also improve the performance and reliability of the AIOps tools that use the data, as the encryption reduces the risk of data corruption, loss, or interference. According to a survey by IDC [4], TLE/SSL encryption can help AIOps tools achieve higher levels of availability, scalability, and efficiency, as the encryption enables faster and more secure data transmission, storage, and analysis.

### E. ENCRYPTION AT REST

One of the best practices to protect the data collected by AIOps systems is to use encryption at rest. Encryption at rest is a technique that encrypts the data before storing it on a disk or a cloud service and decrypts it only when it is accessed by authorized users or applications. Encryption at rest ensures that even if the data storage is compromised, the data remains unreadable and unusable by the attackers. Encryption at rest can also help to comply with the privacy laws and regulations that apply to the sensitive data.

There are two main types of encryption keys that are used for encryption at rest: cloud managed keys and customer-managed keys. Vendor managed keys are the default option, where AIOps vendor handles the encryption and decryption of the data using FIPS 140-2 compliant 256-bit AES encryption [19]. Customer-managed keys, also known as bring your own key (BYOK), offer more flexibility and control to the customers, who can create, rotate, disable, and revoke their own keys using Key Vault. Key Vault is a service that provides secure storage and management of encryption keys, secrets, and certificates. Customer-managed keys can also enable auditing and logging of the key usage and access.

### F. ON-PREMISES DEPLOYMENT

One of the ways to protect your organization's sensitive data is to use on premise deployment for the AIOps system. On premise deployment is a method of hosting and managing the AIOps system on the organization's own servers and infrastructure, rather than using a cloud service provider. On premise deployment can offer more control and security over the data, as the organization can apply its own policies and standards for data access, encryption, backup, and recovery. On premise deployment can also reduce the risk of data breaches or leaks due to external factors, such as network outages, cyberattacks, or legal issues. However, on premise deployment also has some challenges, such as higher upfront and maintenance costs, lower scalability and flexibility, and more dependency on internal IT resources.

### G. ROLE-BASED ACCESS CONTROL

One of the major challenges in AIOps systems is to protect the sensitive data that is collected, processed, and analyzed by the system from unauthorized access or disclosure. Sensitive data can include personal information of the users, business secrets of the organizations, or system configuration and performance data that can reveal vulnerabilities or weaknesses. A common technique to prevent sensitive data exposure is to implement role-based access control (RBAC) for the AIOps system. RBAC is a security model that assigns roles to the users of the system, and grants permissions to the roles based on the principle of least privilege. RBAC also defines policies that specify the conditions and constraints under which the roles can access the data. By using RBAC, the AIOps system can ensure that only the authorized and authenticated users can access the data they need for their tasks, and that they cannot access or modify the data they do not need or are not allowed to. RBAC can also help to reduce the attack surface and the potential impact of data breaches, as well as to comply with the privacy laws and regulations that apply to the sensitive data.

The following three rules are essential for RBAC:
- A subject need to have a role, either by choosing or being assigned one, before they can use a permission.
- A subject's role must be valid and approved.
- A subject can only use a permission that is allowed for their role.

RBAC has many advantages, such as enhancing security and complying with regulations [5]. However, RBAC also has some challenges, such as needing extensive domain expertise, taking a lot of time to implement, and being hard to maintain.

### H. SUPRESS RAW SQL CAPTURE AND QUERY LITERALS

Another way to protect the data collected by AIOps systems is to suppress raw SQL capture and query literals. Raw SQL capture is a feature that collects the SQL statements executed by the application, along with the dynamic parameters bound to runtime values. Query literals are the actual values of the parameters in the SQL statements. Raw SQL capture and query literals can provide useful information for debugging and performance analysis, but they can also expose sensitive data if they contain personal or confidential information. To prevent this, AIOps systems can disable the capture of raw SQL and query literals or mask the values of the sensitive data. Literals that are part of the WHERE clause of an SQL statement are replaced with *****, for example, WHERE userId = '*********'. For example, AppDynamics offers options to disable raw SQL capture, bind variable capture, and query literal capture for its application monitoring solution [6]. Dynatrace also offers similar suppression and disablement of raw SQL [7]. By suppressing raw SQL capture and query literals, AIOps systems can reduce the risk of data exposure and comply with the privacy laws and regulations that apply to the sensitive data.

### I. DATA RETENTION AND ARCHIVING

AIOps tools are dependent on historical performance to produce meaningful insight. It is also important to know how long the data is retained by the AIOps vendors, and how to configure the data retention period according to the needs and preferences of the customers.

Different AIOps vendors have different default data retention periods for the data they collect and analyze. For example, IBM Cloud Pak for Watson AIOps has a default data retention period of 14 days for logs, 15 days for metrics, 90 days for closed alerts, and 30 days for deleted or changed topology resources [8]. AppDynamics has a default data retention period of 4 hours for raw SQL capture and query literals [9]. Dynatrace has a default data retention period of 35 days for metrics, 10 days for logs, and 7 days for user sessions [10]. These default data retention periods can be changed by the customers if they want to keep the data for a longer or shorter time, depending on their requirements and preferences. However, changing the data retention period may have some implications for the performance, scalability, and cost of the AIOps systems.

Therefore, customers should carefully evaluate their data retention needs and options before choosing an AIOps vendor or changing the default data retention period. Customers should also consider the privacy laws and regulations that apply to the sensitive data they collect and store and ensure that they comply with them. By doing so, customers can ensure that they use the data collected by AIOps systems effectively and securely.

The data retention period is a trade-off between accuracy and cost. A shorter period may compromise the quality of the AIOps analysis, while a longer period may incur higher storage expenses and data risks. Therefore, the customer should carefully balance these factors and select the optimal retention period.

### J. SECURITY COMPLIANCE

It is essential to ensure that the data collected by AIOps systems is secure and compliant with the relevant laws and regulations.

Security compliance is the process of adhering to the standards and best practices that aim to protect the data from unauthorized access, disclosure, modification, or destruction. Security compliance can involve various aspects, such as data encryption, access control, auditing, logging, backup, recovery, and incident response. Security compliance can also vary depending on the type, location, and jurisdiction of the data, as well as the industry, sector, and organization that owns or uses the data. Some of the common security compliance frameworks and regulations that apply to the data collected by AIOps systems are:

- Payment Card Industry Data Security Standard (PCI DSS): This is a set of requirements that apply to any organization that processes, stores, or transmits cardholder data, such as credit card or debit card information. PCI DSS aims to ensure the security and privacy of the cardholder data and prevent fraud and identity theft [13].

- ISO 27001: ISO 27001 is an international standard that specifies the requirements for establishing, implementing, maintaining, and improving an information security management system (ISMS). An ISMS is a framework of policies and procedures that includes all legal, physical, and technical controls involved in an organization's information risk management processes. ISO 27001 [14] covers all types of information, regardless of the format, location, or ownership. ISO 27001 certification is a voluntary process that involves an independent audit by an accredited certification body. The audit consists of two stages: Stage 1 is a document review to verify the compliance of the ISMS with the ISO 27001 requirements, and Stage 2 is an on-site visit to validate the implementation and operation of the ISMS. The certification is valid for three years, subject to annual surveillance audits and a recertification audit at the end of the cycle.

- SOC 2 Type 2: SOC 2 Type 2 [15] is a report that provides an independent assessment of the security, availability, processing integrity, confidentiality, and privacy controls of a service organization. A service organization is an entity that provides services to other entities, such as cloud computing, data hosting, software development, or IT outsourcing. SOC 2 Type 2 covers the design and operating effectiveness of the controls over a period of time, typically between six months and one year. SOC 2 Type 2 is based on the criteria set by the American Institute of Certified Public Accountants (AICPA) Trust Services Principles and Criteria. SOC 2 Type 2 attestation is a voluntary process that involves an audit by a licensed Certified Public Accountant (CPA) or accountancy organization. The audit consists of two phases: Phase 1 is a readiness assessment to evaluate the readiness of the service organization to undergo the SOC 2 Type 2 audit, and Phase 2 is the actual SOC 2 Type 2 audit to test the design and operating effectiveness of the controls. The attestation is valid for the period covered by the audit, and can be renewed annually.

- Health Insurance Portability and Accountability Act (HIPAA): This is a US federal law that regulates the privacy and security of protected health information (PHI), which is any information that relates to the health or health care of an individual. HIPAA [16] applies to any organization that creates, receives, maintains, or transmits PHI, such as health care providers, health plans, or health care clearinghouses.

- General Data Protection Regulation (GDPR): This is a European Union (EU) regulation that governs the collection, processing, and transfer of personal data, which is any information that relates to an identified or identifiable individual. GDPR [17] applies to any organization that offers goods or services to individuals in the EU, or monitors the behavior of individuals in the EU, regardless of the location of the organization.

- California Consumer Privacy Act (CCPA): This is a US state law that grants certain rights and protections to the consumers of California regarding their personal information, which is any information that identifies, relates to, describes, or is reasonably capable of being associated with a particular consumer or household. CCPA [18] applies to any organization that does business in California and meets certain criteria, such as having annual gross revenues of more than $25 million, or collecting or selling the personal information of more than 50,000 consumers, households, or devices.

AIOps vendors are the providers of AIOps solutions, such as platforms, tools, or services, that enable the customers to leverage AI and ML for IT operations [11]. AIOps vendors are responsible for ensuring that the data collected by their solutions is secure and compliant with the applicable laws and regulations. AIOps vendors can achieve security compliance by following various steps, such as:

- Conducting a risk assessment and gap analysis to identify the potential threats and vulnerabilities of the data, and the current state and desired state of the security controls and measures.

- Implementing and maintaining the appropriate security controls and measures, such as data encryption, access control, auditing, logging, backup, recovery, and incident response, according to the best practices and standards of the industry and the organization.

- Monitoring and testing the effectiveness and performance of the security controls and measures, and reporting and resolving any issues or incidents that may occur.

- Reviewing and updating the security policies and procedures and providing training and awareness to the staff and the customers, to ensure the alignment and compliance with the changing laws and regulations.

Security compliance is a critical and challenging aspect of AIOps, as it involves the protection and regulation of a large amount of sensitive data that is collected and analyzed by AIOps systems. AIOps vendors play a key role in ensuring that the data collected by their solutions is secure and compliant with the relevant laws and regulations. By doing so, AIOps vendors can enhance the trust and confidence of their customers and deliver value and benefits to their business and IT operations. Customers should also verify that AIOps vendors have the necessary certifications that align with their own security compliance standards.

## V. ACKNOWLEDGMENTS

I would like to thank anonymous reviewers for the comments and suggestions.

## VI. SUMMARY

The article explores the security issues and solutions for AIOps, which applies AI and ML to IT operations. The article highlights that security is a joint duty of the customer and the AIOps vendor.